# Vehicles to Pedestrians Signal Transmissions Based on Cloud Computing

Zhiyi Huang, Junliang Ye, Jiaqi Chen, Xiaohu Ge
School of Electronic Information and Communications
Huazhong University of Science and Technology
Wuhan, China
e-mail: xhge@mail.hust.edu.cn

Yonghui Li
School of Electrical and Information Engineering
University of Sydney
Sydney, Australia
e-mail: yonghui.li@sydney.edu.au

*Abstract*—Collisions between vehicles and pedestrians usually result in the fatality to the vulnerable road users (VRUs). Thus, new technologies are needed to be developed for protecting the VRUs. Based on the high density of pedestrians and limited computing capability of base stations, in this paper the cloud computing technologies are adopted to handle the huge amounts of safety-critical messages. Moreover, the wireless multi-hop backhaul technology is adopted to overcome the bottlenecks of limited transmission capability and queueing delay of the transmitted safety-critical messages between base stations and clouds. Based on the multi-hop wireless transmission scheme, the signal transmission success probability and delay between pedestrians and clouds are derived for performance analysis. Furthermore, numerical simulations are performed to illustrate the relationship between the transmission success probability and the received signal to interference plus noise ratio (SINR) threshold.

*Keywords-cloud computing; delay; multi-hop wireless transmission*

## I. INTRODUCTION

Widely deployed Long Term Evolution based vehicle to everything communication (LTE-based V2X) networks was proposed by 3GPP to support the future self-driving technology in the year 2016. Based on the definition given by 3GPP, the LTE-based V2X networks was constructed with four types of communication methods as vehicle-to-vehicle (V2V) communication, vehicle-to-pedestrian (V2P) communication, vehicle-to-infrastructure (V2I) communication and vehicle-to-network (V2N) communication [1], [2]. The proposed LTE-based V2X networks are aiming at the enhancement of the LTE systems to enable vehicles to communicate with other vehicles, pedestrians and infrastructure in order to exchange the safety-critical messages to ensure road safety, controlling traffic flow, and providing various of intelligent traffic applications [3], [4].

Based on the definition in [5], VRUs are those non-motorized road users, such as pedestrians and cyclists as well as motorcyclists and persons with disabilities or reduced mobility. Based on the statistical results in [6], more than 3000 people worldwide died per day due to the dangerous driving, and half of them are VRUs. The statistical result of traffic accident in China shows that an average of nearly 25,000 pedestrians died in vehicle accidents every year, which accounts for nearly 25% of the total number of deaths during the last ten years [7]. The main cause of traffic accidents is the inability of road users to detect and perceive oncoming dangers before a sufficient amount of time so that reactions for accident avoidance can be taken. Although supporting driving safety is a hot subject in the telecommunications field, the majority of the work has a focus on the V2V and V2I communications for the avoidance of inter-vehicle collisions [8]. VRUs safety solutions can be based on two classes of VRUs detection methods: a) classic sensor based [9] which employs sensors such as RADARs, LASER scanners, IR sensors and imaging sensors (computer vision); b) wireless communication between vehicles and vulnerable road users [10].

In a LTE-based V2X network, there are huge amounts of safety-critical messages need to be transmitted by vehicles, pedestrians and base stations to ensure road safety. However, the computing capability of base stations are limited. Thus, the cloud computing technology should be utilized in LTE-based V2X networks to handle the huge amount of communication messages. And the scalability, resilience, adaptability, connectivity, cost reduction and high performance features of cloud computing technology have high potential to lift the efficiency and quality of information processing [11], [12]. This paper mainly focuses on the success probability of communication and the round trip delay throughout the whole communication procedure among pedestrians, vehicles, base stations and cloud. The main contributions of this paper are three-fold:

1) By utilizing the automatic repeat-request (ARQ) technology as the reliable transmission technology of the network, closed-form formulas for transmission success probability and transmission reliability was derived in this paper. The corresponding simulation results indicate that the number of necessary retransmissions increases significantly with the increase of received SINR threshold;
2) The wireless multi-hop transmission was utilized to overcome the bottleneck of queueing delay of the transmitted safety-critical messages. Based on this configuration, the retransmit delay instead of queueing delay was derived in this paper, The corresponding simulation result shows that the retransmission delay increases rapidly with the increase of received SINR threshold;
3) Based on the proposed system model, the

transmission reliability and delay were derived in this paper. Based on simulation results, we found that the transmission delay and reliability were coupled. When the transmission reliability increases with the received SINR threshold, the transmission delay also increases simultaneously.

The rest of this paper is organized as follows. Section II describes the system model of the whole communication procedure. Network performance metrics are also analyzed and derived in Section II. Numerical and simulation results are displayed in Section III. Finally, Section IV concludes this paper.

## II. SYSTEM MODEL

In the system model of this article, wireless multi-hop technologies are adopted for communication among pedestrians and base stations, vehicles and base stations, base stations and clouds. Based on the fact that the probability of traffic accidents in urban areas with high vehicle density is much higher than that in other areas, a typical urban area is investigated in this paper. The system model in this paper is shown in Fig. 1. Without loss of generality, the typical urban area is configured to be a circular area which includes the vehicles, pedestrians and base stations locate. Assuming that the selected area I is a circular area with a radius $R_b$. Base stations, pedestrians and vehicles are uniformly distributed in this area randomly. Assuming that two-dimensional Cartesian coordinate system xoy is selected, in which the selected circular area satisfies the equation: $x^2 + y^2 \leq R_b^2$. The selected area I includes multiple small areas II with the radius R. In each small area II, there is only one base station, in which the circular area satisfies the equation: $x^2 + y^2 \leq R^2$.

Based on the definition of LTE-based V2X networks, pedestrians and base stations, vehicles and base stations, base stations and cloud have the ability to communicate with each other. By assuming that the working spectrums of the three types of communications are different from each other, there is no cross-layer interference among different types of communications.

The whole communication procedure is divided into the following three parts:
1) Communication between pedestrians and base stations: A device (such as a smart phone) carried by a pedestrian transmits the relevant messages of the pedestrian (such as speed, location, etc.) to a base station;
2) Communication between vehicles and base stations: a) A vehicle transmits its relevant messages (such as speed, location, etc.) to a base station; b) The base station transmits warning messages to the vehicle;
3) Communication between base stations and clouds: a) A base station transmits the messages (such as pedestrians' messages, vehicles' messages, etc.) to cloud; b) The cloud transmits warning messages to the base station.

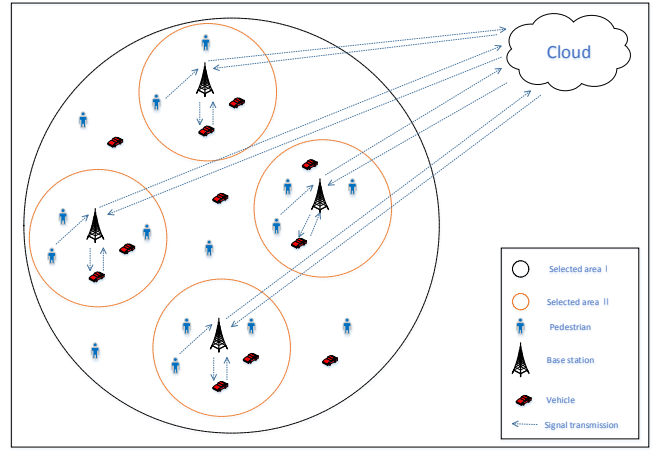

Fig 1. The typical V2X communication scenario in urban area.

## III. VEHICLES TO PEDESTRIANS WIRELESS TRANSMISSIONS

### A. Communication between pedestrians and base stations

Assumed that the location of a base station is $(x_0, y_0)$, the number of pedestrians in the selected area II is M. The location of a pedestrian is $(x_i, y_i)$ $(i=1,2,\cdots,M)$. The distance $d_i$ between the pedestrian and the base station is expressed as

$$d_i = \sqrt{(x_i - x_0)^2 + (y_i - y_0)^2} \quad (i=1,2,\cdots,M), \quad (1)$$

based on the result in [13], the probability density function (PDF) $f_d(x)$ of the distance $d_i$ between the pedestrian and the base station is expressed as

$$f_d(x) = \frac{4x}{\pi R^2} \cdot \left( \arccos \frac{x}{2R} - \frac{x}{2R} \cdot \sqrt{1 - \frac{x^2}{4R^2}} \right), \quad (2)$$

where $0 \leq x \leq 2R$.

The received SINR at the base station from the pedestrian is expressed as

$$SINR_i = \frac{P_t \cdot h_t \cdot d_i^{-\alpha}}{\sigma^2 + \sum_{m=1}^{M_I} \left( P_t \cdot h_m \cdot d_m^{-\alpha} \right)} \quad (i=1,2,\cdots,M), \quad (3)$$

where $M_I$ is the number of interfering pedestrians, and $M_I = M - 1$. $P_t$ is the transmit power of a pedestrian device, $h_t$ is the small scale fading experienced by the desired signal and $h_m$ is the small scale fading experienced by the interfering signal. Based on the assumption that the small scale fading of all channels are independent and identically distributed (i.i.d.) random variables following the Rayleigh distribution, i.e., $h_t$ and $h_m$ are governed by the exponential distribution with parameter λ. $d_i$ is the distance between the i-th pedestrian and the base station, $d_m$ is the distance between the m-th interfering pedestrian and the base station, α is the path loss exponent, $\sigma^2$ is the power of the additive white Gaussian noise (AWGN).

$P_{US}$ is the probability that the base station successfully receives a signal from the pedestrian device, and

$P_{US} = \Pr(SINR \geq \theta)$, where $\Pr(\cdot)$ is the probability that the expression in parentheses is true, and $\theta$ is the threshold of the received SINR at the base station.

By ignoring the shadowing [14], $P_{US}$ is expressed as

$$P_{US} = \Pr\left(\frac{P_t \cdot h_t \cdot d_i^{-\alpha}}{\sigma^2 + \sum_{m=1}^{M_I}(P_t \cdot h_m \cdot d_m^{-\alpha})} \geq \theta\right) \quad (4)$$

$$= \Pr\left(\frac{h_t}{\frac{\sigma^2}{P_t \cdot d_i^{-\alpha}} + \frac{\sum_{m=1}^{M_I} h_m \cdot d_m^{-\alpha}}{d_i^{-\alpha}}} \geq \theta\right).$$

Let $\rho_0 = \frac{P_t \cdot d_i^{-\alpha}}{\sigma^2}$, $\rho_0$ is the received signal to noise ratio (SNR) of a pedestrian device, then (4) is further derived by

$$P_{US} = \Pr\left(\frac{h_t}{\frac{1}{\rho_0} + \frac{\sum_{m=1}^{M_I} h_m \cdot d_m^{-\alpha}}{d_i^{-\alpha}}} \geq \theta\right)$$

$$= \Pr\left(\frac{1}{\frac{1}{\rho_0 \cdot h_t} + \frac{\sum_{m=1}^{M_I} h_m \cdot d_m^{-\alpha}}{h_t \cdot d_i^{-\alpha}}} \geq \theta\right) \quad (5)$$

$$= \Pr\left(\frac{1}{\rho_0 \cdot h_t} + \frac{\sum_{m=1}^{M_I} h_m \cdot d_m^{-\alpha}}{h_t \cdot d_i^{-\alpha}} \leq \frac{1}{\theta}\right).$$

Based on [15], the Laplace transform operation on the random variable Z is denoted by $L_z(s)$, where Z is

$$Z = \frac{1}{\rho_0 \cdot h_t} + \frac{\sum_{m=1}^{M_I} h_m \cdot d_m^{-\alpha}}{h_t \cdot d_i^{-\alpha}}. \quad (6)$$

Thus, the Laplace transform $L_z(s)$ is derived as (7), where $\exp(\cdot)$ is the exponential function and $E_{h_t,h_m,d_i,d_m}(\cdot)$ is the expectation operations on random variables $h_t$, $h_m$, $d_i$ and $d_m$.

$$L_z(s) = E_{h_t,h_m,d_i,d_m}(\exp(-sZ))$$

$$= E_{h_t,h_m,d_i,d_m}\left(\exp(-s(\frac{1}{\rho_0 \cdot h_t} + \frac{\sum_{m=1}^{M_I} h_m \cdot d_m^{-\alpha}}{h_t \cdot d_i^{-\alpha}}))\right). \quad (7)$$

Based on (6) (7), the Laplace transform $L_z(s)$ is derived by

$$L_z(s) = E_{h_t,h_m,d_i,d_m}\left(\exp(\frac{-s}{\rho_0 \cdot h_t}) \cdot \prod_{m=1}^{M_I}\exp(\frac{-s \cdot h_m \cdot d_m^{-\alpha}}{h_t \cdot d_i^{-\alpha}})\right) \quad (8)$$

$$= E_{h_t}\left(\exp(\frac{-s}{\rho_0 \cdot h_t}) \cdot \left(E_{h_m,d_i,d_m}\left(\exp(\frac{-s \cdot h_m \cdot d_m^{-\alpha}}{h_t \cdot d_i^{-\alpha}})\right)\right)^{M_I}\right),$$

and the PDFs of $h_t$, $h_m$, $d_i$ and $d_m$ are

$$\begin{cases} f_{d_i}(x) = \frac{4x}{\pi R^2} \cdot \left(\arccos\frac{x}{2R} - \frac{x}{2R}\sqrt{1-\frac{x^2}{4R^2}}\right) \\ f_{d_m}(x) = \frac{4x}{\pi R^2} \cdot \left(\arccos\frac{x}{2R} - \frac{x}{2R}\sqrt{1-\frac{x^2}{4R^2}}\right) \\ f_{h_t}(x) = \lambda\exp(-\lambda x) \\ f_{h_m}(x) = \lambda\exp(-\lambda x) \end{cases} \quad (9)$$

By substituting (9) into (8), the Laplace transform in (8) becomes (10).

$$L_z(s) = \int_0^\infty \left(\exp(\frac{-s}{\rho_0 \cdot h_t}) \cdot \lambda\exp(-\lambda h_t)\right)dh_t$$

$$\cdot \left(\int_0^\infty \int_0^{2R} \int_0^{2R} \exp(\frac{-s \cdot h_m \cdot d_m^{-\alpha}}{h_t \cdot d_i^{-\alpha}})\right.$$

$$\cdot \frac{4d_i}{\pi R^2} \cdot \left(\arccos\frac{d_i}{2R} - \frac{d_i}{2R}\sqrt{1-\frac{d_i^2}{4R^2}}\right) \quad (10)$$

$$\cdot \frac{4d_m}{\pi R^2} \cdot \left(\arccos\frac{d_m}{2R} - \frac{d_m}{2R}\sqrt{1-\frac{d_m^2}{4R^2}}\right)$$

$$\left.\cdot \lambda\exp(-\lambda h_m)dd_i \cdot dd_m \cdot dh_m\right)^{M_I}.$$

Similar to [16], by the Euler summation on (10), $P_{US}$ in (5) becomes (11).

$$P_{US} = 2^{-B_e} \cdot \theta \cdot \exp(\frac{A_e}{2}) \cdot \sum_{b=0}^{B_e}\binom{B_e}{b} \cdot \sum_{c=0}^{C_e+b}\frac{(-1)^c}{D_c} \cdot \text{Re}\left(\frac{L_z(s)}{s}\right), \quad (11)$$

$$D_c = \begin{cases} 2 & if\ c = 0 \\ 1 & others \end{cases}. \quad (12)$$

According to well established guidelines in [16], [17], the three parameters $A_e$, $B_e$ and $C_e$ control the estimation error, which should be no less than $t\ln 10$, $1.243t-1$ and $1.467t$ respectively to obtain a numerical accuracy of $10^{-t}$. In order to obtain a numerical accuracy of $10^{-8}$, $A_e$, $B_e$ and $C_e$ are configured to be $8\ln 10$, 11 and 14. The parameter $s$ is expressed as $s = \frac{\theta}{2}(A_e + 2\pi cj)$, where $j$ is the imaginary unit. $\text{Re}(\cdot)$ denotes the real part of the given variable in parentheses.

### B. Communication between vehicles and base stations

Assumed that the number of vehicles in the selected area II is N. The location of a vehicle is $(x_j, y_j)(j = 1,2,\cdots,N)$. The number of interfering vehicles is $N_I$, and $N_I = N - 1$.

Assumed that the channel model of the communication between the vehicles and the base station is the same as that between the pedestrians and the base station. The Laplace transform of $L_{z_v}(s)$ can be derived based on Section II.A as

$$L_{z_V}(s) = \int_0^\infty \left( \exp(\frac{-s}{\rho_1 \cdot h_j}) \cdot \lambda \exp(-\lambda h_j) \right) dh_j$$

$$\cdot \left( \int_0^\infty \int_0^{2R} \int_0^{2R} \exp(\frac{-s \cdot h_n \cdot d_n^{-\alpha}}{h_j \cdot d_j^{-\alpha}}) \right.$$

$$\cdot \frac{4 d_j}{\pi R^2} \cdot \left( \arccos \frac{d_j}{2R} - \frac{d_j}{2R} \sqrt{1 - \frac{d_j^2}{4R^2}} \right) \quad (13)$$

$$\cdot \frac{4 d_n}{\pi R^2} \cdot \left( \arccos \frac{d_n}{2R} - \frac{d_n}{2R} \sqrt{1 - \frac{d_n^2}{4R^2}} \right)$$

$$\left. \cdot \lambda \exp(-\lambda h_n) dd_j \cdot dd_n \cdot dh_n \right)^{N_I},$$

where $\rho_1 = \frac{P_e \cdot d_j^{-\alpha}}{\sigma^2}$, $P_e$ is the transmit power of a vehicle signal. $h_j$ is the small scale fading experienced by the desired signal and $h_n$ is the small scale fading experienced by the interfering signal. $d_j$ is the distance between the j-th vehicle and the base station, $d_n$ is the distance between the n-th interfering vehicle and the base station.

Based on (11), (12) and (13), the probability that the base station successfully receives the signal transmitted by the vehicle is $P_{VS}$, and

$$P_{VS} = 2^{-B_e} \cdot \theta \cdot \exp(\frac{A_e}{2}) \cdot \sum_{b=0}^{B_e} \binom{B_e}{b} \cdot \sum_{c=0}^{C_e+b} \frac{(-1)^c}{D_c} \cdot \text{Re}\left( \frac{L_{z_V}(s)}{s} \right). \quad (14)$$

### C. Communication between base stations and clouds

Assumed that the location of the cloud is $(x_c, y_c)$, the number of base station in the selected area I is K. The location of a base station is $(x_k, y_k)(k=1,2,\cdots,K)$. The number of interfering base stations is $K_I$, and $K_I = K - 1$.

Assumed that the channel model of the communication between the base stations and the cloud is the same as that between the pedestrians and the base station. The Laplace transform of $L_{z_B}(s)$ can be derived based on Section II.A as

$$L_{z_B}(s) = \int_0^\infty \left( \exp(\frac{-s}{\rho_2 \cdot h_k}) \cdot \lambda \exp(-\lambda h_k) \right) dh_k$$

$$\cdot \left( \int_0^\infty \int_0^{2R_b} \int_0^{2R_b} \exp(\frac{-s \cdot h_r \cdot d_r^{-\alpha}}{h_k \cdot d_k^{-\alpha}}) \right.$$

$$\cdot \frac{4 d_k}{\pi R_b^2} \cdot \left( \arccos \frac{d_k}{2R_b} - \frac{d_k}{2R_b} \sqrt{1 - \frac{d_k^2}{4R_b^2}} \right) \quad (15)$$

$$\cdot \frac{4 d_r}{\pi R_b^2} \cdot \left( \arccos \frac{d_r}{2R_b} - \frac{d_r}{2R_b} \sqrt{1 - \frac{d_r^2}{4R_b^2}} \right)$$

$$\left. \cdot \lambda \exp(-\lambda h_r) dd_k \cdot dd_r \cdot dh_r \right)^{K_I},$$

where $\rho_2 = \frac{P_b \cdot d_k^{-\alpha}}{\sigma^2}$, $P_b$ is the transmit power of a base station signal. $h_k$ is the small scale fading experienced by the desired signal and $h_r$ is the small scale fading experienced by the interfering signal. $d_k$ is the distance between the k-th base station and the cloud, $d_r$ is the distance between the r-th interfering base station and the cloud.

Based on (11), (12) and (15), the probability that cloud successfully receives the signal transmitted by the base station is $P_{BS}$, and

$$P_{BS} = 2^{-B_e} \cdot \theta \cdot \exp(\frac{A_e}{2}) \cdot \sum_{b=0}^{B_e} \binom{B_e}{b} \cdot \sum_{c=0}^{C_e+b} \frac{(-1)^c}{D_c} \cdot \text{Re}\left( \frac{L_{z_B}(s)}{s} \right). \quad (16)$$

### D. Communication delay

Considering wireless communication scenarios are limited in urban areas, the wireless propagation distance is short and the wireless propagation delay can be ignored compared with the signal processing delay in pedestrians, vehicles and base stations. Thus, we mainly focus on signal processing time.

*1) Communication delay between pedestrians and base stations:* The probability that a base station successfully receives a signal transmitted by a pedestrian device is $P_{US}$. In order to ensure that a pedestrian device signal is successfully received by a base station, the number of the transmissions from the pedestrian to the base station is $M_U$, and $M_U = 1/P_{US}$.

The length of a signal data and the processing rate of a pedestrian device are $l_U$ and $\mu_U$, respectively. The time for a pedestrian device to process a signal data is $t_U$, and $t_U = l_U / \mu_U$.

The time for a pedestrian device to successfully transmits a signal is $T_U$, and

$$T_U = t_U \cdot M_U. \quad (17)$$

*2) Communication delay between vehicles and base stations:* The probability that a base station successfully receives a signal transmitted by a vehicle is $P_{VS}$. In order to ensure that a vehicle signal is successfully received by a base station, the number of the transmissions from the vehicle to the base station is $N_V$, and $N_V = 1/P_{VS}$.

The length of a signal data and the processing rate of a vehicle are $l_V$ and $\mu_V$, respectively. The time for a vehicle to process a signal data is $t_V$, and $t_V = l_V / \mu_V$.

The time for a vehicle to successfully transmits a signal is $T_V$, and

$$T_V = t_V \cdot N_V. \quad (18)$$

*3) Communication delay between base stations and clouds:* The probability that cloud successfully receives a signal transmitted by a base station is $P_{BS}$. In order to ensure that a signal of a base station is successfully received by the cloud, the number of the transmissions from the base station to the cloud is $N_{BS}$, and $N_{BS} = 1/P_{BS}$.

The length of a signal data and the processing rate of a base station are $l_{BS}$ and $\mu_{BS}$, respectively. The time for a base station to process a signal data is $t_{BS}$, and $t_{BS} = l_{BS} / \mu_{BS}$.

The time for a pedestrian device to successfully transmits a signal is $T_{BS}$, and

$$T_{BS} = t_{BS} \cdot M_{BS}. \quad (19)$$

*4) The whole communication procedure:* Based on the above analysis, the success probability and the total delay of signals transmission throughout the whole communication procedure are $P_S$ and $T$, respectively, and

$$P_S = P_{US} \cdot P_{VS} \cdot P_{BS}, \quad (20)$$

$$T = T_1 + T_{BS} \quad (T_1 = \max(T_U, T_V)). \tag{21}$$

## IV. NUMERICAL SIMULATION RESULTS

Based on the system model in this paper, the numerical results are shown in this section. Table I summarizes parameters used in this paper [18]-[20].

TABLE I. SYSTEM RELATED PARAMETERS

| Parameter | Description | Value |
|---|---|---|
| $R_b$ | Radius of the selected area I | 500m |
| R | Radius of the selected area II | 200m |
| σ | The power of the additive white Gaussian noise (AWGN) | 1 |
| λ | The parameter of the exponential distribution | 1 |
| $l_U$ | Pedestrian device signal length | 5KB |
| $μ_U$ | Pedestrian device signal processing rate | 2GB/S |
| $l_V$ | Vehicle signal length | 5KB |
| $μ_V$ | Vehicle signal processing rate | 4GB/S |
| $l_{BS}$ | Base Station signal length | 10KB |
| $μ_{BS}$ | Base Station signal processing rate | 8GB/S |

This paper focuses on the probabilities of successful transmission of pedestrian signals, vehicle signals and base station signals, and the number of transmissions required for successful signal transmission under different SINR thresholds. Experimental numerical results are as follows:

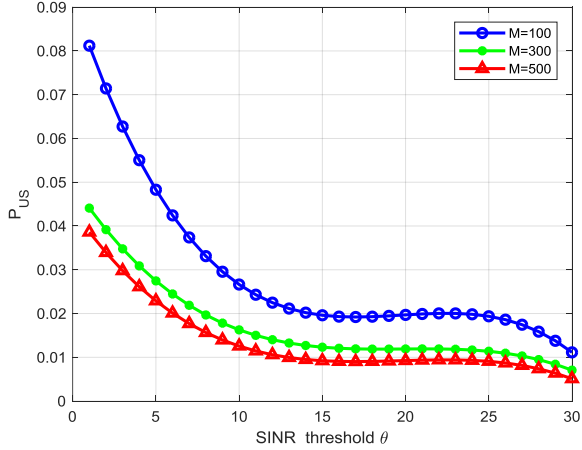

Fig 2. Pedestrian signal transmission success probability $P_{US}$ with respect to the SINR threshold θ under different numbers of pedestrians.

The impact of the SINR threshold θ and the numbers of pedestrians M on the pedestrian signal transmission success probability $P_{US}$ is investigated in Fig. 2. When θ is fixed, $P_{US}$ decreases with the increase of M. When M is fixed, $P_{US}$ decreases with the increase of θ.

Based on Section II.B, the impact of the SINR threshold θ and the number of vehicles N on the vehicle signal transmission success probability $P_{VS}$ is same as the curve trend investigated in Fig. 2. When θ is fixed, $P_{VS}$ decreases with the increase of N. When N is fixed, $P_{VS}$ decreases with the increase of θ.

Based on Section II.C, the impact of the SINR threshold θ and the number of base stations K on the base station signal transmission success probability $P_{BS}$ is same as the curve trend investigated in Fig. 2. When θ is fixed, $P_{BS}$ decreases with the increase of K. When K is fixed, $P_{BS}$ decreases with the increase of θ.

The numbers of pedestrians, vehicles and base situations of the three situations are as follows: A: M=100, N=50, K=3; B: M=300, N=100, K=5; C: M=500, N=150, K=7. The impact of the SINR threshold θ on the signal transmission success probability $P_S$ and the total communication delay T under three situations of A, B and C are investigated in Fig. 3 and Fig. 4, respectively. When θ is fixed, $P_S$ decreases and T increases under the three situations of A, B and C. When M, N and K are fixed, $P_S$ decreases and T increases with the increase of θ under the three situations of A, B and C.

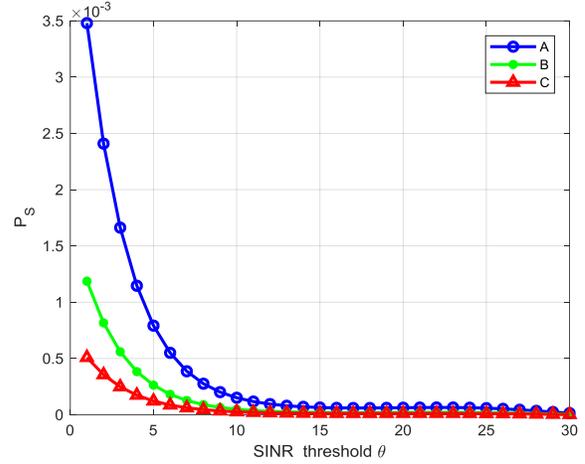

Fig 3. Signal transmission success probability $P_S$ with respect to the SINR threshold θ under different three situations throughout the whole communication procedure.

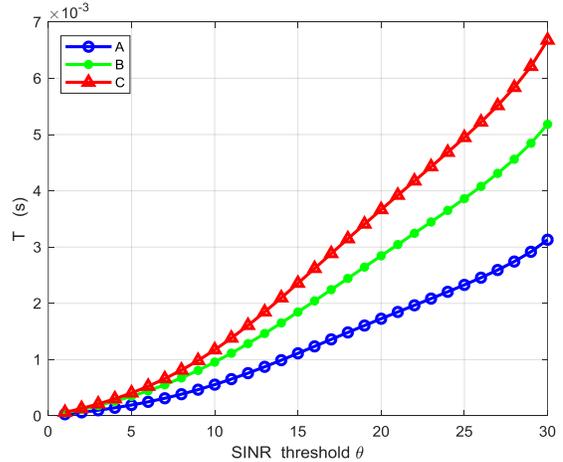

Fig 4. Total communication delay T with respect to the SINR threshold θ under different three situations throughout the whole communication procedure.

## V. CONCLUSION

In this paper, we mainly focus on the analysis of transmission reliability and delay of safety-critical messages in V2X networks. By modeling the locations of pedestrians and base stations as randomly, the distances and aggregated interference among pedestrians, base stations and vehicles are analyzed. Moreover, the wireless multi-hop technology is adopted in this paper to overcome the bottleneck of limited transmission capability and queueing delay of the transmitted safety-critical messages between base stations and clouds, and the closed-form expressions of transmission success probability, transmission reliability and transmission delay are derived. By doing simulations on the proposed system model, we found that the transmission delay and reliability were coupled, and when transmission reliability increases with the received SINR threshold the transmission delay also increases simultaneously.

ACKNOWLEDGMENT

Authors thank the support from National Key R&D Program of China (2016YFE0133000): EU-China study on IoT and 5G.